\documentclass[structabstract,referee]{aa}

\usepackage{natbib}

\usepackage{graphicx}

\def\ion[#1 #2]{#1\,{\sc #2}}
\def\ergs[#1]{#1 {ergs}~{cm$^{-2}$}\,{s$^{-1}$}\,{sr$^{-1}$}}
\def\dens[#1]{10$^{#1}$\hskip 1.5pt{cm$^{-3}$}}
\def\densr[#1 #2]{10$^{#1}$\hskip 1pt{--}\hskip .5pt{10$^{#2}$}\hskip 1.5pt{cm$^{-3}$}}
\def\fl[#1 #2]{{#1}$\pm${#2}}
\def\orb[#1 #2]{{$#1^{#2}$}}
\def\ls[#1 #2]{{$^{#1}${#2}}}
\def\tm[#1 #2 #3]{{$^{#1}${#2}$_{#3}$}}

\begin{document}

\title{The role of radiative losses in the late evolution of pulse-heated coronal loops/strands}

\author{F. Reale\inst{1}\inst{2} \and E. Landi\inst{3}}

\institute{Dipartimento di Fisica, Universit\`a di Palermo, Piazza del Parlamento 1, 90134, Italy\\
\and INAF-Osservatorio Astronomico di Palermo, Piazza del Parlamento 1, 90134 Palermo, Italy\\
\and Department of Atmospheric, Oceanic and Space Sciences, University of Michigan, Ann Arbor, MI 48109}

\abstract{Radiative losses from optically thin plasma are an important ingredient for modeling 
plasma confined in the solar corona. Spectral models are continuously updated to include the 
emission from more spectral lines, {with significant effects on radiative losses, especially 
around 1~MK}.}
{We investigate the effect of changing the radiative losses temperature dependence due to 
upgrading of spectral codes on predictions obtained from modeling plasma confined in the 
solar corona.}
{The hydrodynamic simulation of a pulse-heated loop strand is revisited comparing 
results using {an old} and {a recent} radiative {losses} function.}
{We find significant changes in the plasma evolution during the late {phases} of plasma 
cooling: {when} the {recent radiative loss curve is used}, the plasma {cooling rate increases significantly
when temperatures reach 1-2~MK}. {Such more rapid cooling occurs when the plasma density is larger than a threshold 
value}, and therefore {in impulsive heating models that cause the loop plasma 
to become overdense. The fast cooling has the effect of steepening the slope of
the emission measure distribution of coronal plasmas with temperature at temperatures lower 
than $\sim 2$~MK.}}
{{The effects} of {changes in the} radiative losses {curves} can be important 
for modeling the late phases of the evolution of pulse-heated coronal loops, and, more in general, of thermally unstable optically thin plasmas.}


\authorrunning{Reale \& Landi}
\titlerunning{Importance of radiative losses for coronal loops}

\maketitle

\section{Introduction}
\label{sec:intro}

Coronal loops are the building blocks of the solar corona. They
consist of pipe-like magnetic field structures arching in the corona 
and connecting photospheric magnetic regions of opposite polarity.
Loops are filled with optically thin, hot and relatively dense plasma,
with temperatures ranging from 0.8~MK to a few million degrees, depending
on the regions where they are located. There are two main classes of
mechanisms that have been proposed to explain the heating of the 
loop plasma to coronal temperatures: steady \citep[or "high frequency",][]{Warren2011a} heating mechanisms, and 
impulsive (or "low frequency") heating mechanisms. Loop models that include steady
heating predict a steady-state plasma, while in the pulse-heated
scenario the plasma evolves dynamically and spends most of the time in a 
cooling state \citep{Cargill1994a}. In particular, a fast energy pulse 
might heat the plasma to more than 10 MK for a few seconds 
\citep[e.g.,][]{Cargill1994a,Cargill2004a,Reale2008a,Guarrasi2010a}; then,
the plasma is free to cool down losing energy by conduction to the chromosphere 
and by radiation. The balance between these two loss mechanisms is 
essentially determined by the local density: during and immediately
after the heat pulse the loop plasma density is still low and conduction 
dominates, then denser plasma from the chromosphere fills the loop
and radiation becomes the most efficient loss mechanism
\citep{Antiochos1980a,Cargill2004a,Reale2007b,Reale2010a}. Observational evidence seems to
indicate that
loop plasma might be heated impulsively \citep[as reviewed in][]{Klimchuk2006a},
and recent studies using Hinode and Solar Dynamics Observatory data seem to confirm it
\citep{Guarrasi2010a,Reale2011a,Terzo2011a,Viall2011a}. 

Regardless of the heating scenario, radiative losses are an 
important mechanism of plasma cooling in coronal loops and need 
to be included with accuracy in loop models \citep[e.g.,][]{McClymont1983a,Antiochos1982a,
Bradshaw2003a,Muller2003a,Bradshaw2005a}. They become particularly 
important if loops are described as bundles of thin strands \citep{Cargill1993a,Klimchuk2006a}, 
each ignited by a single short and intense heat pulse \citep{Parker1988a}
as the plasma cooling times are far longer than the expected duration 
of the heat pulse so that each strand spends most of its time cooling 
by radiation. Radiative losses are calculated by summing the emission 
of the plasma over the entire wavelength spectrum. The radiative emission 
of the optically thin plasma is dominated by the bremsstrahlung and 
free-bound recombination continua, and by line emission from all ions 
of all elements present in the coronal plasma. The total radiative 
losses are approximately proportional to the square of the electron 
density of the plasma \citep{Tucker1966a,Landini1970a,Tucker1971a};
\citet{Landi1999a} showed that departures from such proportionality 
are lower than 25\%. The dependence on the electron temperature, on 
the contrary, is much more complex and, in typical coronal conditions, 
can be parameterized with a separate function of the temperature 
obtained by fitting a known curve to the total radiative losses of 
isothermal spectra calculated using a grid of temperature values 
\citep{Rosner1978a,Landi1999a}.

The isothermal spectra used to determine the total radiative loss curve
are obtained using a spectral code, which collects all the relevant atomic
parameters and transition rates necessary to calculate the line and continuum
emission of an optically thin plasma. The most popular codes available in the
literature are CHIANTI \citep{Dere1997a,Landi2012a}, AtomDB {\citep{Foster2012a}},
and SPEX \citep{Kaastra1996a}; as new and improved calculations of atomic data
and transition rates become available in the literature, these codes are 
costantly updated to extend their calculations to new lines neglected before, 
or improve the results for {those already available}. As a consequence,
predicted {radiative losses} change, and differences from those calculated with 
earlier versions of the codes can sometimes be large.

The consequences of upgrades to spectral codes and radiative losses can be 
important both for data analysis and diagnostics \citep{Testa2012a}, and 
for plasma modeling \citep{Soler2012a}. For example, time-dependent hydrodynamic 
plasma models have been extensively used to describe loop plasma evolution. 
All models include the radiative losses in the energy balance, sometimes
with a temperature dependence described with a piecewise power law function 
as in \cite{Rosner1978a}, which is based on the \cite{Raymond1977a} 
spectral model. More recent calculations of the radiative losses have
a much larger loss rate at around 1~MK, due to higher metal abundances and 
the inclusion of a large amount of spectral lines from \ion[Fe viii-xv] as 
well as other elements formed at similar temperatures. {Differences at
higher temperatures are more limited.} Thus, for a plasma impulsively heated 
to about 10 MK, we do not expect large differences for most of the evolution. 
However, when the plasma cools down and approaches 1~MK, the much larger loss 
rate will cause the loop to cool more efficiently.


In this paper we will show in detail that, in a pulse-heated loop model, the 
enhanced radiative losses more easily and earlier lead to catastrophic cooling, 
that qualitatively changes the plasma evolution and may have important implications 
on both the observed emission distribution in active and quiet regions, and on 
the loop heating mechanisms as well.  Our approach is to revisit a well-tested 
multi-strand pulse-heated loop model \citep{Reale2008a,Guarrasi2010a} using the 
standard and an updated radiative losses function and discuss the differences 
in the results.

%
%

Section~\ref{sec:model} summarizes the loop model focusing on the radiative losses 
function, Section~\ref{sec:results} illustrates the results using different functions, Section~\ref{sec:diagn} describes the implications for diagnostics, 
 Section~\ref{sec:disc} discusses the results and their implications. 

\section{The model}
\label{sec:model}

We consider a hydrodynamic simulation identical to the one in \citet{Guarrasi2010a} 
that models a pulse-heated loop strand with the Palermo-Harvard loop code 
\citep{Peres1982a,Betta1997a}. The loop {strand} is semicircular, symmetric 
with respect to the apex, and its symmetry axis is perpendicular to the solar 
surface. The loop half-length is $L = 3 \times 10^9$ cm and it includes a 
chromospheric layer at the footpoints that is linked to the corona through 
a steep transition region. 

The plasma confined in each strand transports energy and moves only along the 
magnetic field lines, and its evolution can be described with a one-dimensional 
time-dependent hydrodynamic model \citep[e.g.,][]{Nagai1980a, Peres1982a, 
Doschek1982a, Nagai1984a, McClymont1983a, MacNeice1986a, Gan1991a, Hansteen1993a, 
Betta1997a, Antiochos1999a, Muller2003a, Bradshaw2003a, Bradshaw2006a}, through 
the equations \citep{Peres1982a,Betta1997a}:

\begin{equation}
	\frac{d n}{d t} ~ = -n \frac{\partial \textrm{v}}{\partial s}
\end{equation}

\begin{equation}
	n m_H \frac{d \textrm{v}}{d t} ~ = -\frac{\partial p}{\partial s} + n m_H g + \frac{\partial }{\partial s} \left( \mu  \frac{\partial \textrm{v}}{\partial s} \right) 
\end{equation}

\begin{equation}
	\frac{d \varepsilon}{d t} + w \frac{\partial \textrm{v}}{\partial s} ~ = Q - n^2 \beta P(T) + \mu \left( \frac{\partial \textrm{v}}{\partial s} \right)^{2} + \frac{\partial }{\partial s} \left(  \kappa T^{5/2} \frac{\partial T}{\partial s} \right) 
\end{equation}

\begin{equation}
	p ~ = \left( 1+\beta \right) n k_B T
\end{equation}

\begin{equation}
	\varepsilon ~ = \frac{3}{2} p + n \beta \chi
\end{equation}

\begin{equation}
	w ~ = \frac{5}{2} p + n \beta \chi
\end{equation}

where $n$ is the hydrogen number density; $t$ is time, $s$ is the 
field line coordinate; $\textrm{v}$ is the plasma velocity; $m_H$ is 
the mass of hydrogen atom; $p$ is the pressure; $g$ is the component 
of gravity along the field line; $\mu$ is the effective coefficient 
of compressional viscosity (including numerical viscosity); 
$\beta = n_e / n$ is the ionization fraction where $n_e$ is the 
electron density; $T$ is the temperature; $\kappa$ is the thermal 
conductivity ($\simeq 9 \times 10^{-7}$ erg~cm$^{-1}$s$^{-1}$K$^{-7/2}$); 
$k_B$ is the Boltzmann constant; $\chi$ is the hydrogen ionization 
potential; $P\left( T \right) $ is the radiative losses function per 
unit emission measure (discussed later); $Q \left( s,t \right) $ is the power input per 
unit volume:

\begin{equation}
	Q \left( s,t \right) ~ = H_0 + H_1 f\left( t \right)  
\end{equation}
$H_{0}$ is a low-regime ($H_0 = 3 \times 10^{-5}$ erg cm$^{-3}$ s$^{-1}$) 
steady heating term which balances radiative and conductive losses for the 
static initial atmosphere. {$H_1$ is the amplitude of} the heat pulse, 
and is assumed to be uniform along the loop. The time dependence of the heat pulse is a top-hat 
function, with $f(t) = 1$ for $0 < t < 60$s and $f(t) = 0$ at any other 
time. The amplitude of the pulse is $H_1 = 0.38$ erg cm$^{-3}$ s$^{-1}$. We 
have checked that the presence of the steady heating term ($H_0$) is irrelevant 
for the entire strand evolution.
For further investigation, we will also show simulations with different 
intensity of the heat pulse.

The initial condition is that of a very low-pressure loop atmosphere, 
with a base pressure of $p_0 \approx 0.055$ dyne~cm$^{-2}$, which results in 
an apex temperature of $T_0 \approx 8.0 \times 10^{5}$K. The chromosphere is 
assumed to be model F in \citet{Vernazza1981a}, and its energy 
balance is strictly maintained at all times.

The Palermo-Harvard loop code \citep{Peres1982a,Betta1997a} has been 
extensively used  for to model both flaring \citep{Peres1987a,Betta2001a} 
and quiescent loops \citep{Reale2000a,Guarrasi2010a}. The code has an adaptive mesh 
refinement \citep{Betta1997a}, to achieve adequately high resolution 
in the steep gradients along the strand and during the evolution.

\begin{figure}
\includegraphics[width=8.0cm]{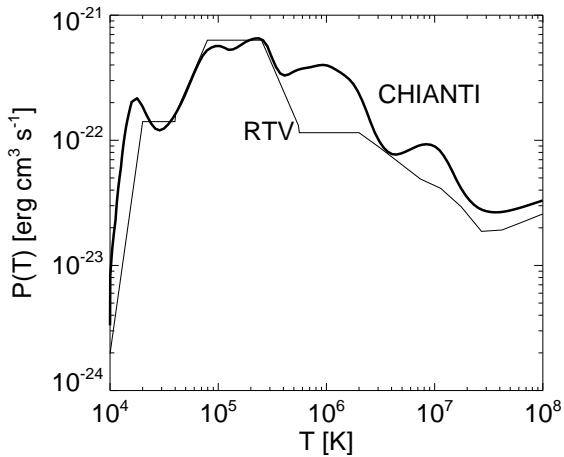}
\caption{Radiative losses function $P(T)$ (emissivity per unit emission 
measure) as a function of temperature according to \citet{Rosner1978a} 
({\it thin solid line}) and to {version 7} of CHIANTI spectral 
code \citep{Landi2012a} ({\it thick solid line}).}
\label{fig:pt}
\end{figure}

We replicate the simulation of \citet{Guarrasi2010a} with two different 
radiative losses functions $P(T)$, shown in Figure~\ref{fig:pt}: one from 
\citet{Rosner1978a} (hereafter RTV) computed according to \citet{Raymond1977a},
the other computed according to version 7 of the CHIANTI code 
\citep{Landi2012a}, {assuming a density of $10^9$ cm$^{-3}$ and ionization equilibrium according to \citet{Dere2009a}}. {We have checked that radiative losses functions adopted in other works, e.g. \citet{Klimchuk2008a}, fall in the range between these two curves and therefore we expect intermediate results when using them.}

The differences between the two curves are due to three sources. First, 
the atomic models in CHIANTI are much larger and more sophisticated than 
those available to \citet{Raymond1977a},  and such larger atomic models
provide vast amounts of additional spectral lines.
The main differences fall around {0.5-3}~MK, because of the increase in 
size of the \ion[Fe viii] to \ion[Fe xiv] models, and around 10~MK,
because of the larger models for \ion[Fe xviii-xxiii]. Increases in 
the size of the models for ions of other elements have a more limited 
effect. Overall, \citet{Landi1999a} found that increases in CHIANTI 
version~2 atomic models and improvements on the atomic data caused 
radiative losses to change by as much as 50\%. As even {bigger} 
models are used in CHIANTI version~7, we expect differences to be 
{larger}.

The second source of variation lies in the ion fractions used in the two
calculations. Ionization and recombination rates used to calculate the
charge state composition at equilibrium are normally taken from theoretical
calculations, which have been dramatically improved during the 30-40 years
that separate the \citet{Raymond1977a} and {the latest version of the}
CHIANTI code. The differences due to ion fraction improvements are expected 
to affect all elements and ions. \citet{Landi1999a}, for example, estimated 
that different ion abundance datasets caused the radiative losses to change 
by up to 40\%. 

The third
source of variation are element abundances. These are expected to provide
large effects: \citet{Landi1999a} reported differences of factors up
to 2.5 when different abundance datasets are used. In the present comparison,
the abundances used by \citet{Raymond1977a} are the cosmic values reported
by \citet{Allen1973a}, while the CHIANTI code radiative losses were calculated
using the coronal abundances by \citet{Feldman1992a}. The main difference 
between the two abundance datasets lies in the fact that the cosmic abundances 
of elements with First Ionization Potential (FIP) smaller than 10~eV have been 
increased in the coronal abundance dataset by a factor $\approx$3.5 to account 
for the element fractionation in the solar corona known as the FIP effect 
\citep{Feldman1992b}; additional, much smaller differences are found between 
the photospheric element abundances of ions with FIP$>$10~eV and the cosmic 
values. The factor 3.5 enhancement involves most of the major elements emitting 
in the corona -- Mg, Si and Fe -- so that large effects are expected at 
temperatures in the 1-15~MK, where Fe emission dominates the spectrum.

\section{The results}
\label{sec:results}

\begin{figure}
\includegraphics[width=8.0cm]{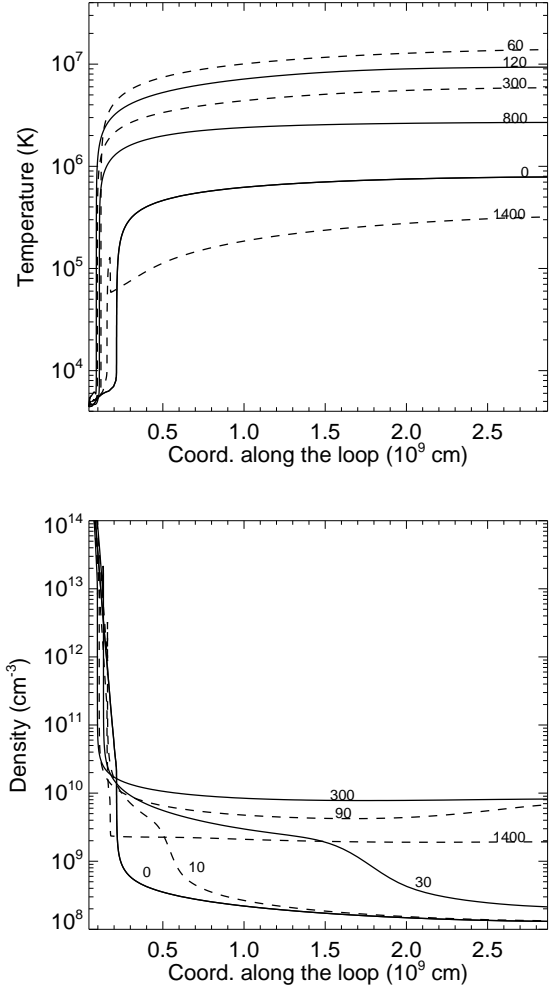}
\caption{Model temperature (top) and density (bottom) along half of the pulse-heated 
{loop strand. The loop apex is at the right end of the X-axis. The plasma temperature
and densities are displayed at several different times (in s) as reported in the
two panels, from 0 (heat pulse start) to 1400s. }}
\label{fig:evol}
\end{figure}

We have calculated the evolution of the loop-confined plasma for $\sim 2000$~s after the start of the impulsive heating. Such an evolution is 
known from  many previous studies \citep[e.g.,][]{Peres1993a,Warren2002a,Warren2003a,Patsourakos2005a,Reale2008a,Guarrasi2010a}. Figure~\ref{fig:evol} shows samples of the temperature and 
density profiles along half of the strand at several different times
(0s, 10s, 30s, 60s, 90s, 120s, 300s, 800s and 1400s), that {summarize} the 
entire {strand} evolution. { Figure~\ref{fig:evol} displays} the simulation 
computed with the CHIANTI radiative losses, but the evolution obtained {using 
the other loss curve} is overall very similar. {Because of} the strong heating 
pulse, the plasma rapidly heats to above 10~MK (60s) along 
most of the loop. Then it gradually and uniformly cools down, {and reaches} 
$\sim 0.1$~MK, i.e. {below} {the} initial temperature, {in} about half 
an hour. The density follows a slightly different evolution and with a different 
timing. It increases initially with an evaporation front coming up from the 
chromosphere ($t=10$s, 30s). After the front has reached the loop apex, the 
density increases more uniformly and reaches its maximum after about 5 minutes, 
i.e. much later than the end of the heat pulse. Then also the density begins to 
decrease, due to draining driven by the cooling \citep{Bradshaw2010a}. After about half an hour, the 
density is still much higher than it was before the heat pulse.  

\begin{figure}
\includegraphics[width=10.0cm]{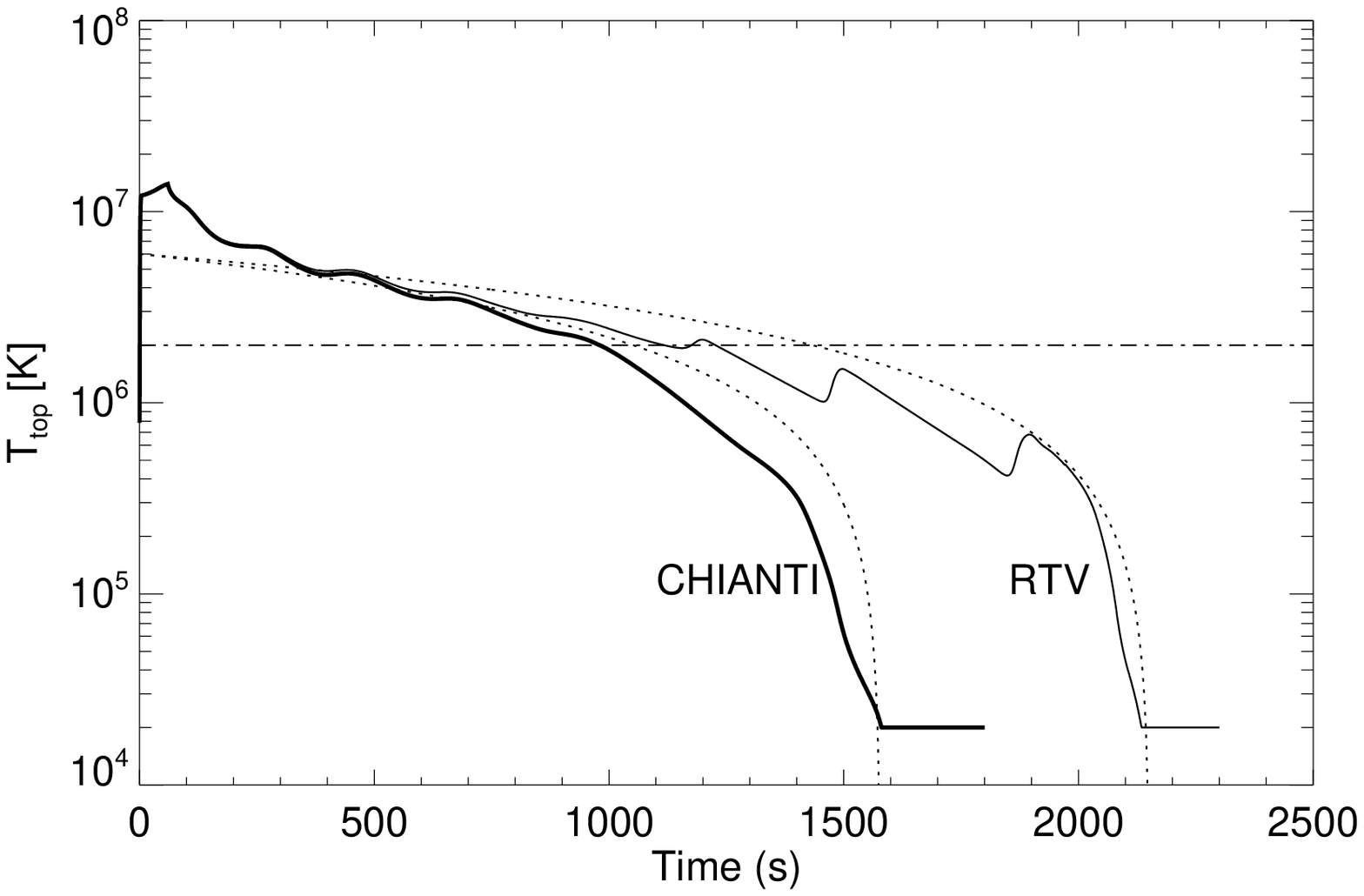}
\includegraphics[width=10.0cm]{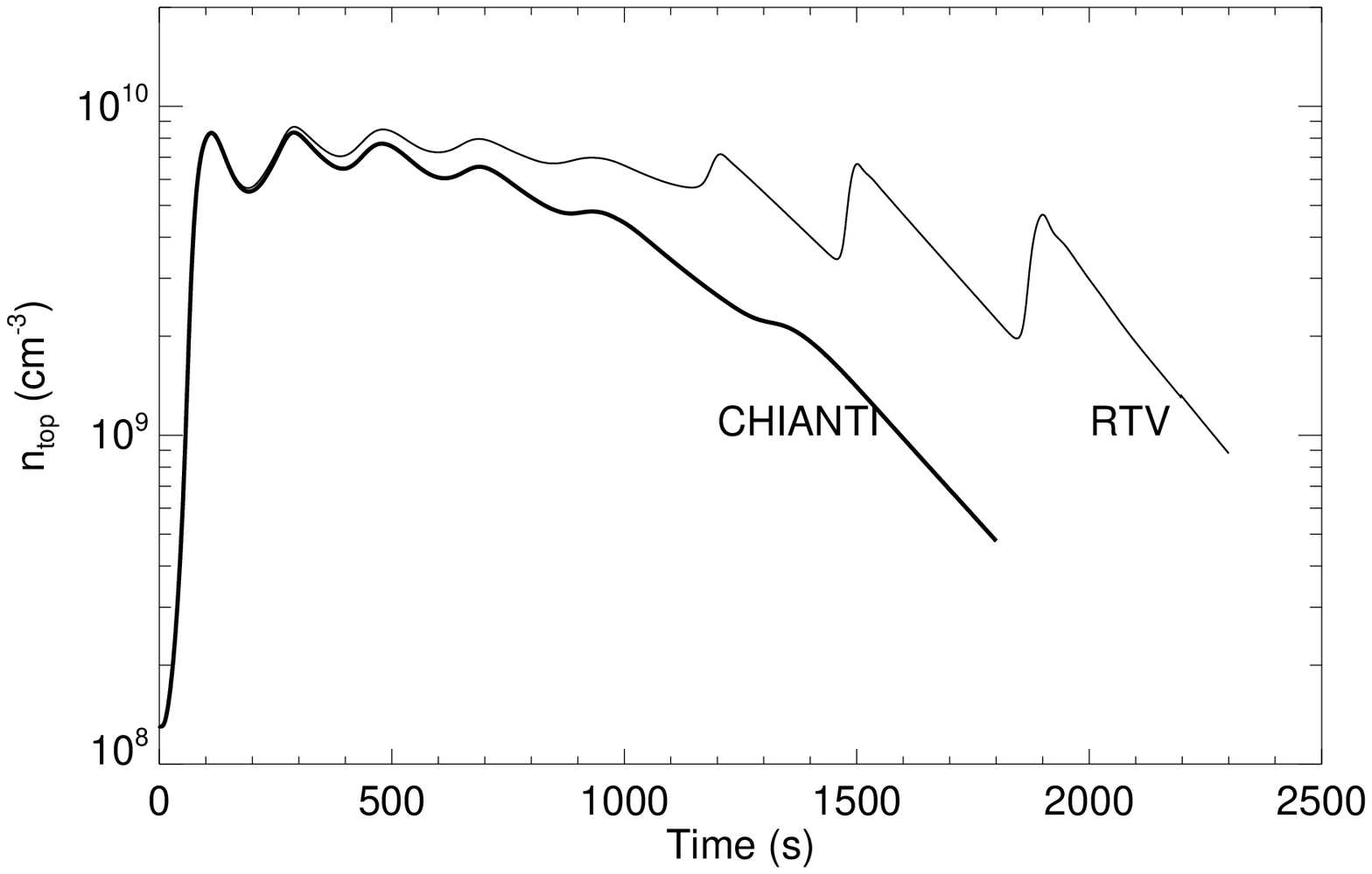}
\caption{Evolution of the temperature (top) and density (bottom) at the loop 
apex from hydrodynamic simulations using the two different radiative {loss functions:} \cite{Rosner1978a} 
({\it thin solid line}) and {version 7 of CHIANTI} ({\it thick solid line}).  Results obtained with analytical approximations \citep{Cargill1994a} are also shown ({\it dotted lines}, see text for details). The level of 2 MK is marked ({\it dash-dotted horizontal line}).}
\label{fig:top2}
\end{figure}

The effect of the different radiative loss functions is more significant at 
relatively late times. {Such difference is illustrated in Figure~\ref{fig:top2},} that shows the evolution of the temperature and density at the loop apex {obtained
using the two different loss functions}. Both temperature and 
density at the loop apex {are the same during} the first ten minutes 
of evolution, {when the temperature is larger than $\simeq$2~MK. After 10
minutes,} the {evolutionary tracks diverge}: the plasma temperature and density 
{calculated with the CHIANTI radiative losses} decrease faster.  

The temperature evolution ends up with catastrophic cooling 
\citep{Parker1953a,Field1965a} in both cases with temperature dropping abruptly 
by almost one order of magnitude to the minimum of $\sim 2 \times 10^4$ K in 
little more than 3 minutes. Analytical approximations of the rate of decrease 
of the plasma temperature have been derived in the past 
\citep{Antiochos1980a,Cargill1994a,Cargill1995a} and can be compared to those 
obtained numerically. Figure~\ref{fig:top2} shows two solutions obtained using 
the analytical expression \citep{Cargill1994a}:

\begin{equation}
T_{top}(t)=T_0 \left[ 1 - \frac{3}{2} \left( \frac{1}{2} - \alpha \right) \frac{t}{\tau_r}\right] ^{1/(1/2-\alpha)}
\label{eq:carg}
\end{equation}
where
\begin{equation}
\tau_r = \frac{3 k_B T_0}{n_0 P(T)_0}
\label{eq:taur}
\end{equation}
For both cases, we have assumed $n_0 = 7 \times 10^9$ cm$^{-3}$ as density value 
, $T_0 = 6$ MK as starting temperature of the 
radiative phase, {from inspection of the simulations (see Fig.~\ref{fig:top2})}, and $\alpha = -0.5$ as effective power-law index of the radiative 
losses function approximated with $P(T) = \chi T^{\alpha}$ in the temperature range 
of interest, i.e. $5.5 < \log T < 7$ \citep{Cargill1994a}. {We have checked that the chosen values of $n_0$ and $T_0$ are in agreement with those  
obtained from equating the thermal conductive and radiative cooling timescales \citep{Cargill2004a}.} The two
solutions {that match well those of the simulations are obtained using two different values of the effective radiative losses, i.e. 
$P(T)_0 = 1.5 \times 10^{-22}$ and $1.1 \times 10^{-22}$ erg~cm$^3$~s$^{-1}$ 
(see Fig.~\ref{fig:pt}). Substituting in Eq.(\ref{eq:taur})} we obtain $\tau_r \approx 2400$s and 3200s, for 
the faster (CHIANTI) and slower (RTV) cooling, respectively. The decay after the 
initial impulsive evolution, i.e. as soon as the temperature settles to about 6~MK, 
is generally well-described by Equation~(\ref{eq:carg}), although some details are 
lost. In particular, the catastrophic cooling that occurs at $\sim 0.3$~MK is 
well-reproduced. The numerical solution with CHIANTI predicts a faster decay 
below $\sim 2$~MK, that leads to catastrophic cooling at $t \sim 1400$ s, i.e. $\sim 600$ s earlier than with RTV losses. This approximately corresponds to {the} two 
important changes of slope in the CHIANTI radiative loss function 
(Figure~\ref{fig:pt}) at $T \sim 3$~MK  and 0.5~MK. {Therefore, simulations are in good agreement with analytical descriptions, in which we change only the "effectiveness" of the radiative losses in a certain temperature range, i.e. average values of P(T). }

The density decreases more gradually than the temperature \citep{Bradshaw2010a}. 
In the RTV solution we notice saw-toothed behaviour due to plasma bouncing back 
and forth along the closed loop during draining. For our purposes, it is more 
important to look at the average evolution. On average, the density obtained 
with CHIANTI losses decreases considerably faster than with RTV losses. {The basic reason is that the more effective cooling makes pressure decrease faster in absolute value, and the pressure gradient, i.e. the force that sustains the plasma against draining by gravity, decreases faster as well, causing a faster draining. At this point the process is highly non-linear, i.e. catastrophic, and cannot be stopped even if the radiative cooling rate decreases with the density. Bradshaw and Cargill (2010) show the effect in terms of enthalpy, i.e. a peak of radiative losses is followed by a sharp increase of enthalpy losses.}

Overall, 
we might say that the density stays almost steady for more than 1500s with RTV 
losses,  while its reduction becomes significant already after about 1000s with 
CHIANTI losses. This qualitative difference has important implications for the 
emission measure distributions.   

\begin{figure}
\includegraphics[width=10.0cm]{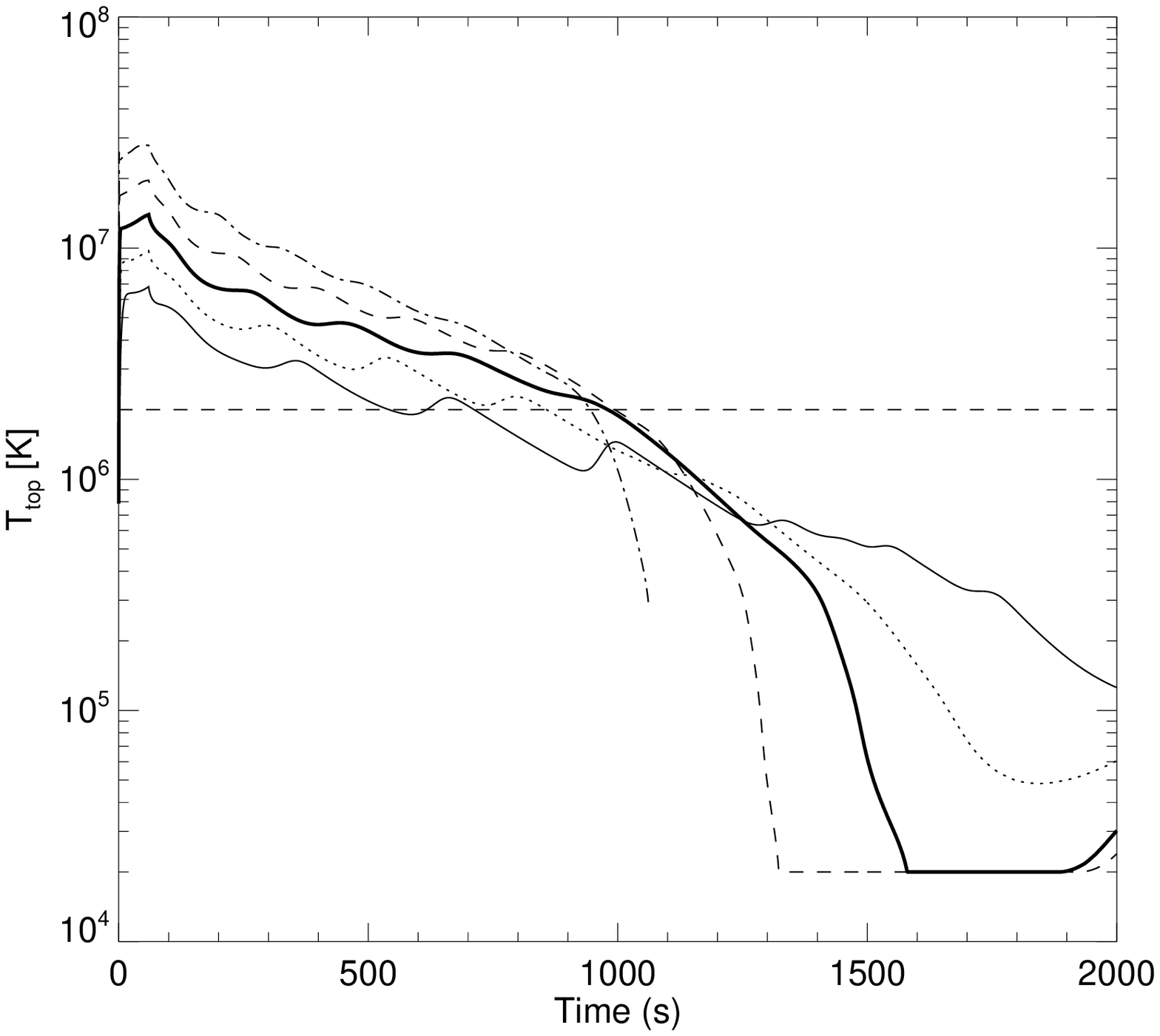}
\includegraphics[width=10.0cm]{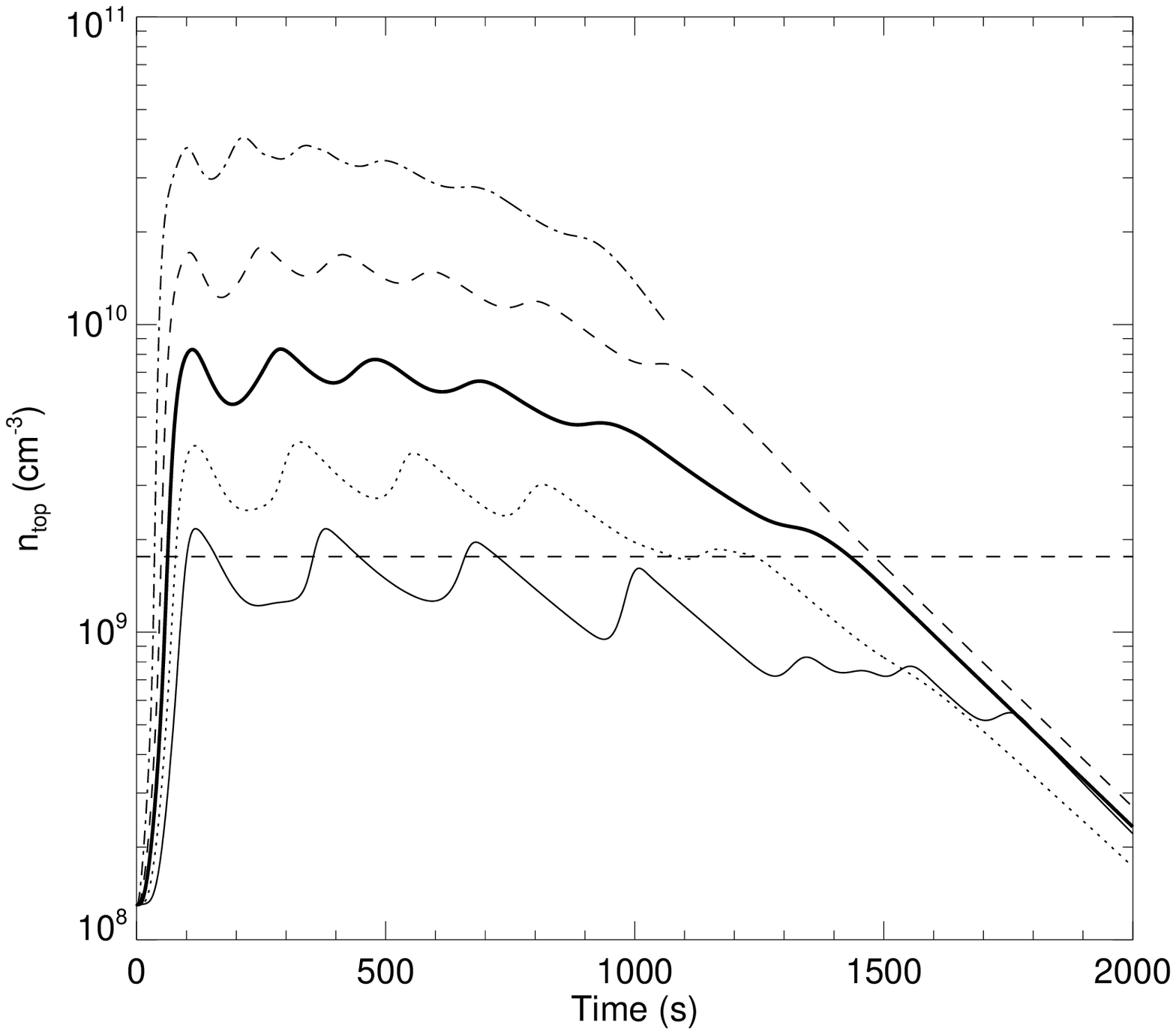}
\caption{Evolution of the temperature (top) and density (bottom) at the loop 
apex obtained with the CHIANTI radiative losses and different {values of 
the} heat pulse { intensity}. The density equilibrium value ({\it dashed 
horizontal line, bottom panel}) for a loop at $\sim 2$ MK ({\it dashed 
horizontal line, top panel}) is marked.}
\label{fig:topm}
\end{figure}

A very important qualitative difference of using CHIANTI instead of RTV  
radiative losses is {the switching from regimes where there is a transition to catastrophic cooling to regimes where the rate of cooling remains roughly constant, i.e.  both radiation and enthalpy losses increase smoothly with time, as shown in Bradshaw and Cargill (2010). This corresponds to} the change of slope of the cooling occurring at about 2~MK. We have 
investigated {in more detail what are the conditions required to trigger this change of behaviour}.
Figure~\ref{fig:topm} shows the evolution of the temperature and density at the 
loop apex obtained with the CHIANTI radiative losses and different {values of
the} heat pulse {intensity $H_1$}. {Higher $H_1$ values lead both to higher 
apex maximum temperature and density values.}  Figure~\ref{fig:topm} {shows
that} the change of slope has a strong dependence on the density, 
{regardless of the heat pulse intensity. In fact, the slope does not 
change if the loop plasma density remains approximately below the equilibrium 
density of a loop at a maximum temperature of about 2~MK.} According to 
\citet{Rosner1978a} this density threshold can be obtained as:

\begin{equation}
n_9 \approx \frac{5}{L_9}
\label{eq:dens}
\end{equation}

\noindent
where $n_9$ is the density in units of $10^9$ cm$^{-3}$ and $L_9$ is the loop 
half-length in units of $10^9$ cm. {If the density is lower than this threshold value then the loop does not enter a catastrophic cooling phase.}

\section{Diagnostic implications of different radiative loss curves}
\label{sec:diagn}

{The} change of cooling rate has {two} important 
{consequences. First,} an important diagnostic {result}  {that can be} 
obtained from coronal observations is the distribution of the emission measure 
as a function of temperature (hereafter EM(T)), 

\begin{equation}
   EM(T) = \int_{0}^{V(T)}{n(T)^2 {dV}}
\label{eq:demt}
\end{equation}

\noindent
Typical coronal {EM(T)} distributions monotonically increase by a few orders 
of magnitude from $\sim 0.1$ to $\sim 1$~MK, have a broad peak around 2-3~MK and 
then decrease again to higher temperature, more steeply than they rise 
\citep[e.g.,][]{Peres2000a,Testa2011a,Warren2011a}. 
The EM(T) curve of a coronal loop can be predicted from theoretical models of loop strands 
like the one we used here by assuming that the loop consists of many strands, each heated randomly in time.
Thus, the overall EM(T) curve will be determined by assuming that at any given time
there will be loop strands at all stages of the strand evolution, and the total
EM(T) curve can be obtained by averaging the EM(T) curve of a single strand over
the entire strand evolution, in this case 2000~s.

\begin{figure}
\includegraphics[width=8.0cm]{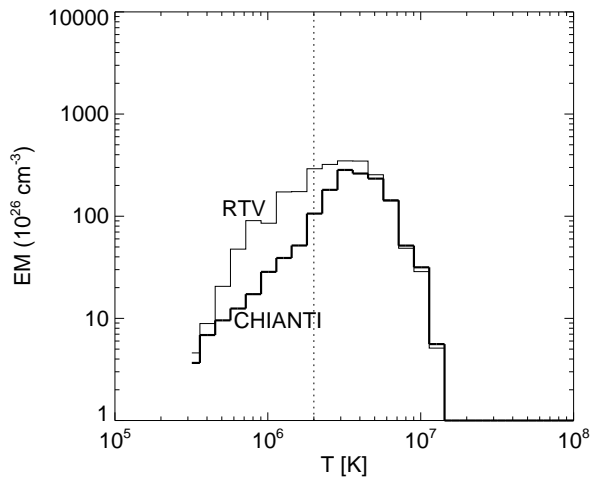}
\caption{{EM(T) curves obtained averaging the emission measure distribution vs 
temperature over the first 2000s of strand evolution,} obtained from the simulation 
with the RTV ({\it thin solid line}) and the Chianti ({\it thick solid line}) 
radiative losses. The temperature of 2 MK is marked for reference ({\it dotted vertical line}).}
\label{fig:emt2}
\end{figure}

In Figure~\ref{fig:emt2} we plot the space- and time-{averaged} EM(T) curves obtained from the 
simulations {made using} the RTV and the CHIANTI radiative losses, in the coronal part of the strand, i.e. excluding moss regions \citep{Guarrasi2010a,Warren2011a}.
The figure shows that the high temperature portion of the EM(T) 
curve does not change in the two calculations, confirming that using different
radiative loss curves does not change the results above 3-4~MK. On the contrary, 
the low temperature part is significantly different, due to the different slopes of the cooling at late times: the RTV simulation leads to a slightly higher peak and to a shallower distribution on the low temperature 
tail, that steepens below $\sim 0.5$ MK. On the other hand, the faster cooling obtained with 
CHIANTI leads to a steeper curve {below 2~MK, and thus to a sharper peak
at 3~MK.} The {absolute} difference {between the two curves reaches} 
a factor 5 around 1~MK; {however,} the difference in the slope {of the 
two curves} is even more important because {such slope} has been {found
to be} an important parameter to discriminate between low frequency and high frequency coronal heating 
mechanisms \citep[e.g.][]{Warren2011a}. In the temperature range 
$5.9 \leq \log T \leq 6.5$ the slope in logarithmic scale is $\approx 1.1$ for 
{the RTV curve}, and $\approx 2.0$ for {the CHIANTI curve}. 
{These values are smaller than those obtained by \citet{Warren2011a}, but one cannot exclude that they may be compatible when all possible uncertainties in data analysis and issues regarding DEM reconstruction \citep{Testa2011a} are taken into account.}

\begin{figure}
\includegraphics[width=8.0cm]{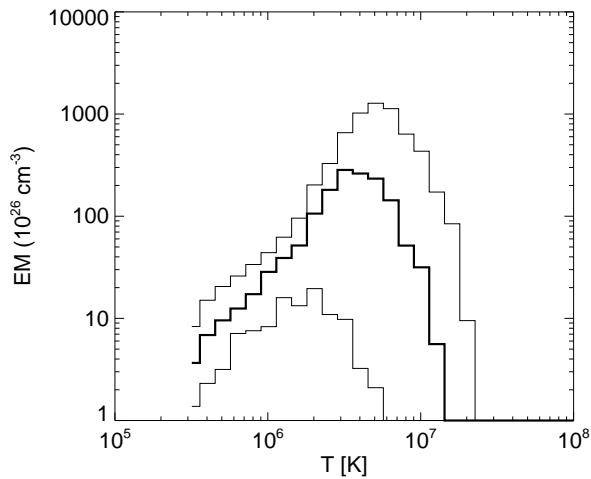}
\caption{{EM(T) curves obtained averaging the emission measure distribution vs
temperature over the first 2000s of strand evolution, obtained from three simulations
made using CHIANTI radiative losses and three different heat pulse amplitudes:
the same used in Figure~\ref{fig:emt2} ({\it thick solid line}), a lower one}
({\it lower thin solid line}), {and a larger one} ({\it upper thin solid line}) (see also Fig.~\ref{fig:topm}).}
\label{fig:emtv}
\end{figure}

Moreover, 
the two different {slopes} on the cool side {of the EM(T) curves} lead 
to different {intensity ratios of EUV lines frequently observed by the
available instrumentation} \citep{Warren2011a}. Figure~\ref{fig:emtv} shows 
two more distributions obtained with CHIANTI radiative losses and different heat 
pulse intensities, {one with a smaller
$H_1$ value leading to a} density steadily below the threshold value in Eq.~(\ref{eq:dens}), 
{one with a larger $H_1$ leading to a} very high density (Fig.~\ref{fig:topm}). The figure 
clearly {shows} that the trend on the cool side of the EM(T) becomes shallower 
{and more similar to the RTV results} at low density: {this means that i)} 
the different slope is linked to the presence of different cooling rates, {and ii) 
it can in principle be used to efficiently discriminate between coronal heating rates 
and mechanisms}.

%
%

\begin{figure}
\includegraphics[width=10.0cm]{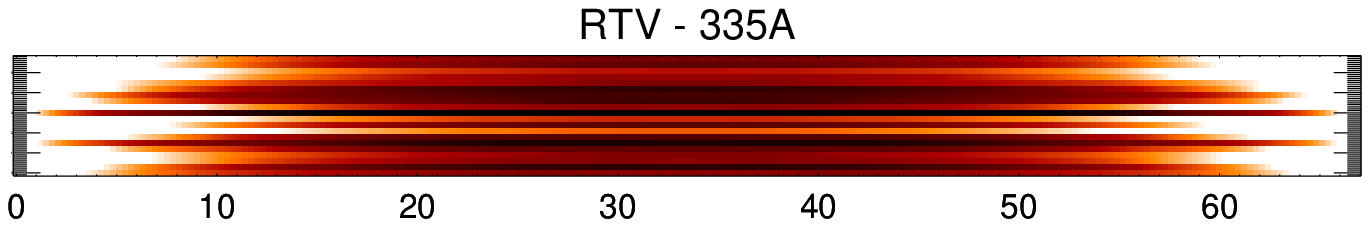}
\includegraphics[width=10.0cm]{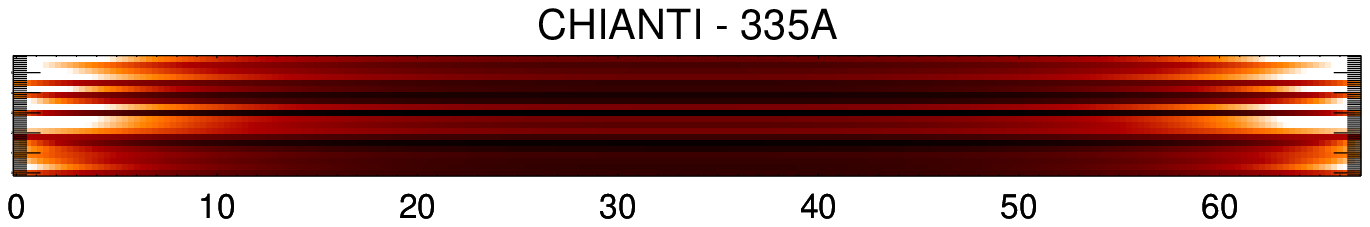}
\includegraphics[width=10.0cm]{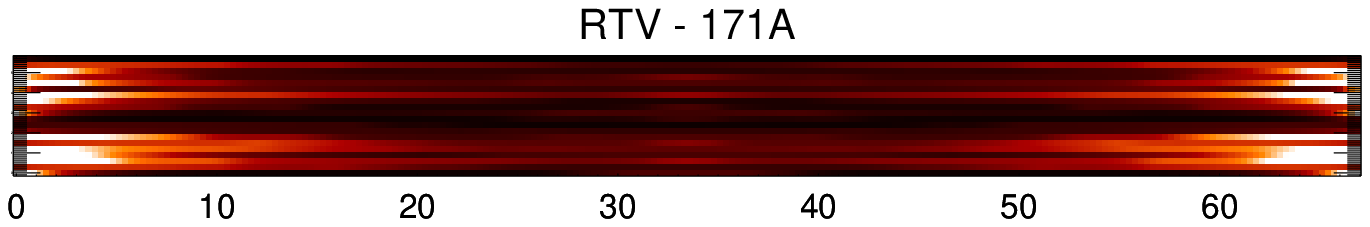}
\includegraphics[width=10.0cm]{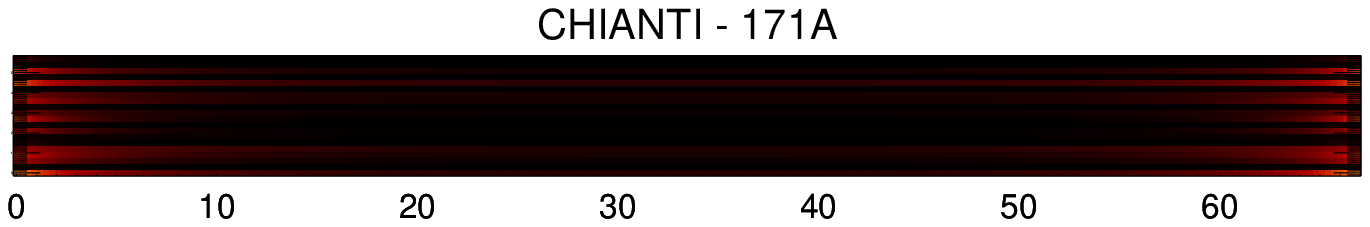}
\caption{Synthetic (straightened) loop images obtained from collecting groups of 10 randomly heated strands, in SDO/AIA 335\AA\ and 171\AA\ channels, with models using RTV and CHIANTI radiative losses. The color scale is linear and exactly the same for a given channel (white is bright). The X-axis is measured in arcsec. }
\label{fig:im}
\end{figure}

The faster cooling with CHIANTI losses below 2~MK has {a second,} important 
implication. {The plasma in the strand} spends a {much longer} time at temperatures 
well above 1-2 MK than below them. If a loop is made by a 
multitude of strands, each heated once {randomly}, {most of the strands within 
a loop will have a high temperatures at any given time, while only few of them 
can be observed} at temperatures at or below $\sim 1$ MK. This might explain, 
at least in part, why active regions {have been found to be} mostly covered by hot 
$\sim$3~MK loops {with} X-ray images {and} SDO/AIA 335\AA\ channel images, and much 
less {populated} by warm $\sim$1~MK loops, {like in the EUV 171\AA\ channel in} 
TRACE or SDO/AIA images, i.e. few loops are seen cooling from hot to warm status, 
much fewer than expected. To illustrate this effect, Figure~\ref{fig:im} shows 
synthetic images of a multi-stranded loop in the two SDO channels obtained from 
the hydrodynamic simulations. We consider 200 straight strands grouped in bundles 
of 10 to mimic a limited instrument resolution. The straight aspect may recall 
a loop projected on the solar disk. Each strand has the evolution described above, 
but the start time of the heat pulse is shuffled: the strands brighten at random 
times. We take a snapshot in a regime situation, i.e. the emission spatially 
averaged over the stranded loop is equal to the average emission over the whole 
strand time evolution. We show images obtained both with RTV and CHIANTI radiative 
losses. The color scale is linear and chosen to be exactly the same in each channel. 
In the 335\AA\ channel we see differences in the details but overall the loop appears 
to be bright. In the 171\AA\ channel, there is a big qualitative change: the loop is 
much fainter, almost invisible, with the CHIANTI radiative losses. {In a nutshell,
the CHIANTI radiative losses significantly decrease the time that a given loop strand
spends at a temperature between 1 and 3~MK, so that on the average it will be more
difficult to observe loops in many narrow-band channels from AIA, TRACE, STEREO/EUVI and
SOHO/EIT than expected when using RTV radiative losses. Since the onset of the 
accelerated (and then catastrophic) cooling depends on the plasma density, and hence
on the rate of impulsive heating, changes in the radiative losses can have significant
effects on heating rates determined from narrow-band images. }

\section{Discussion and conclusions}
\label{sec:disc}

In this paper we have studied the effects of changes in the radiative loss curve
on the evolution of an impulsively heated loop strand. Radiative losses can change
due to 1) improvements in atomic models with the inclusion of more accurate atomic
data and transition rates, or lines previously unavailable; 2) upgrades in the plasma
ion abundance composition; 3) changes in the plasma element abundance composition.
We have carried out our tests using two radiative loss curves, whose differences were
due to all three these causes. The most recent curve, from CHIANTI version 7, had
much more efficient radiative losses at temperatures in the 0.5-3~MK range than the
older one (RTV losses). We expect similar effects also using other losses curves obtained from other spectral codes, with similar atomic models and element abundances (Sec.\ref{sec:intro}). More efficient radiative losses may have effects on modeling other thermally unstable optically thin plasmas, such as flares, supernova remnants, accretion columns from circumstellar disks, novae, galactic cooling 
flows. Here we focus on {their} effects on the physics and structure of coronal loops, 
more in particular on the evolution of plasma confined inside a coronal magnetic flux 
tube and subject to a {short} and strong heat pulse.

The effect is very small in the initial phases of the evolution. Most previous studies 
of the flaring plasma {focused} on the rise and initial decay phase of the flare, 
and much less on the very late phases, {so their results are not affected by changes
in the radiative losses. On the contrary, \citet{Reale2012a} studied the EUV late phase
of a flare using the older RTV radiative losses, and were able to reproduce the observed
light curves with success. The agreement they found between model and observations 
suggests two possibilities. First,} a small change of model parameters may be enough 
to maintain the same degree of agreement {if the CHIANTI} radiative loss function
{is used: for example,} increased radiative losses may require a {smaller} 
heating to lead to the final {catastrophic cooling} that explained the observations. 
{Second,} the element abundances {in the flaring plasma may be lower} than 
{the coronal values assumed in the CHIANTI rates used in the present work so that 
the differences between the RTV and the CHIANTI radiative losses will be smaller.}

{Our results show that changes in the radiative loss curves} have important effects 
{on} the late phases of a pulse-heated coronal loop strand, because they can change 
the cooling rate of the heated plasma and on the timing of the final catastrophic 
cooling. This can lead to immediate and interesting consequences on what we can 
expect from observations. 

{First,} the change of {cooling rate} occurs at relatively low 
temperature ($\sim 2$ MK), and leads to considerable steepening of the cool side 
of the emission measure distribution. {This change in EM(T) slope has important 
effects on our understanding of coronal plasma heating. In fact, {earlier} 
studies of pulse-heated loop models, {using RTV-like radiative losses} provided relatively flat EM(T) curves and 
such broad curves, not observed in active regions, have been} invoked as an 
argument against the effectiveness of {impulsive} heating in {non-flaring} 
active regions \citep{Warren2011a}. {{Thus,} the change in slope caused 
by an improved radiative loss function} can lead to a significant change of 
perspective. 

Second, the temperature of the cooling plasma {decreases} more rapidly below 
2~MK, and this might explain why we see very few loops cooling from X-rays to 
EUV bands in active regions.

{Third,} it is also interesting to notice that we do {not} expect faster cooling
to occur in lower-regime loops{: if the loop is heated by less intense pulses or {in a more gradual} and longer-lasting way, the plasma might never start a fast cooling.} 
Whenever we observe 1 MK loops in UV ``warm'' channels {this might {provide} 
an upper limit to the intensity and duration of the heat pulses that provide 
the energy}.

{Fourth, \citet{Reale2012a} showed that a catastrophic cooling occurred at
the end of the evolution of a post-flare loop, and explained the so-called
EUV late phase observed by EVE during the decay of flares \citep{Woods2011a}. 
Changes in radiative
losses cause the predictions of the EUV late phase of flares to change considerably,
and affect any estimate of flare impulsive heating rates made combining cooling
loop models with observed EUV light curves.}

This work uses one {the latest release} of an up-to-date spectral model. However, 
{chances are that spectral codes will keep evolving and include many more lines and 
more accurate data as they become available}. For example, there is evidence that we 
may still miss many {important spectral lines} \citep{Testa2012a} {that can
have significant effects on the radiative losses. As a consequence,} we expect even 
more significant effects due to further spectral model improvements, so the present 
work {can even be considered} conservative {to this respect. Also, element
abundances are very important for radiative losses, so studies that provide better 
constraints on abundances are highly encouraged. On the contrary, effects of
deviations from ionization equilibrium of the emitting species are negligible in the
late phase of the flare,} because the plasma is still 
at relatively high density, {so that} the ionization equilibrium times are quite 
short {and the plasma is able to respond quickly and adapt to the rapid changes 
in the temperature} \citep{Reale2012a}.

In conclusion, in this work we report on the effect of upgrading the spectral models 
of optically thin emitting plasmas on modeling coronal plasma. In particular, we find 
that, {while} the increased emissivity obtained with new models has negligible 
effects on most of the plasma evolution predicted by models and in many plasma 
conditions, there is a very important implication for modeling of plasma confined 
in coronal loop strands heated by strong and fast energy pulses. Enhanced plasma 
emission {in the 0.5-3~MK temperature range} leads to faster cooling that might explain several pieces of evidence that questioned the validity of pulse-heated loop models. We point out that this kind of effects might be important also for modeling other thermally unstable optically thin plasmas, such as supernova remnants, accretion columns from circumstellar disks, novae.
%

\bigskip

We thank Peter Cargill for useful suggestions and the anonymous referee for constructive comments.
Fabio Reale acknowledges support from Italian Ministero dell'Universit\`a e Ricerca and 
from Agenzia Spaziale Italiana (ASI), ASI/INAF agreement I/023/09/0. The work of Enrico Landi is 
supported by NASA grants NNX10AQ58G and NNX11AC20G, and by NSF grant AGS-1154443. 

\bibliographystyle{aa}
\bibliography{refs_cite}



\end{document}